\documentclass
[aps,pra,amsfonts,amssymb,twocolumn,amsmath,preprintnumbers,nofootinbib,floatfix,superscriptaddress,longbibliography]{revtex4-1}%
\usepackage[dvips]{graphics}
\usepackage{graphicx}
\usepackage{bm}
\usepackage{amsmath}
\usepackage{amsfonts}
\usepackage{amssymb}
\usepackage{xcolor}
\usepackage{subfigure}
\usepackage{tabularx}
\usepackage{multirow}
\usepackage{braket}
\usepackage{commath}
\usepackage[colorlinks,linkcolor=blue,anchorcolor=blue,citecolor=blue,urlcolor=blue]%
{hyperref}%
\providecommand{\U}[1]{\protect\rule{.1in}{.1in}}
\newcommand{\bCalG}{\boldsymbol{\mathcal{G}}}
\begin{document}

\title{Dynamical Chiral Nernst Effect in Twisted Van der Waals Few Layers}

\author{Juncheng Li}
\thanks{These authors contributed equally to this work.}
\affiliation{New Cornerstone Science Laboratory, Department of Physics, The University of Hong Kong, Hong Kong, China}
\affiliation{HKU-UCAS Joint Institute of Theoretical and Computational Physics at Hong Kong, Hong Kong, China}

\author{Dawei Zhai}
\thanks{These authors contributed equally to this work.}
\affiliation{New Cornerstone Science Laboratory, Department of Physics, The University of Hong Kong, Hong Kong, China}
\affiliation{HKU-UCAS Joint Institute of Theoretical and Computational Physics at Hong Kong, Hong Kong, China}

\author{Cong Xiao}
\email{congxiao@um.edu.mo}
\affiliation{Institute of Applied Physics and Materials Engineering, University of Macau, Taipa, Macau, China}
\affiliation{HKU-UCAS Joint Institute of Theoretical and Computational Physics at Hong Kong, Hong Kong, China}

\author{Wang Yao}
\email{wangyao@hku.hk}
\affiliation{New Cornerstone Science Laboratory, Department of Physics, The University of Hong Kong, Hong Kong, China}
\affiliation{HKU-UCAS Joint Institute of Theoretical and Computational Physics at Hong Kong, Hong Kong, China}

\begin{abstract}
	
The Nernst effect is a fundamental thermoelectric conversion phenomenon that was deemed to be possible only in systems with magnetic field or magnetization. In this work, we propose a novel dynamical chiral Nernst effect that can appear in two-dimensional van der Waals materials with chiral structural symmetry in the absence of any magnetic degree of freedom. This unconventional effect is triggered by time variation of an out-of-plane electric field, and has an intrinsic quantum geometric origin linked to not only the intralayer center-of-mass motion but also the interlayer coherence of electronic states. We demonstrate the effect in twisted homobilayer and homotrilayer transition metal dichalcogenides, where the strong twisted interlayer coupling leads to sizable intrinsic Nernst conductivities well within the experimental capacity. This work suggests a new route for electric control of thermoelectric conversion.

\end{abstract}
\maketitle

\section{Introduction}\label{sec1}

The Nernst effect, in which an electric Hall current is generated in response to an applied temperature gradient,
is not only a central ingredient for energy-harvesting thermoelectric devices, but also a fundamental physical effect for probing quantum geometric properties of solids.
While the ordinary Nernst effect is triggered by external magnetic field, the anomalous Nernst effect originates from the momentum-space Berry curvature \cite{Ong2004}, an intrinsic quantum geometric quantity related to the center-of-mass motion of Bloch electrons \cite{Xiao2006,Xiao2010}. Such an intrinsic Nernst response eliminates the necessity of applying a magnetic field, and has attracted extensive experimental studies in recent years \cite{Behnia2017,ikhlas2017large,Sakai2018,Guin2019}.
On the other hand, the requirement for magnetic ordering raises the issue of stability of thermoelectric materials against external magnetic perturbations, as well as excludes nonmagnetic materials from the playground of intrinsic Nernst response.

In this work we propose an alternative scheme for triggering intrinsic Nernst response by electrical means without invoking any magnetic degrees of freedom.
Our proposal is based on the gate tunability of two-dimensional (2D) van der Waals (vdW) layered materials and novel quantum geometric properties of layered electrons, whose wave functions are distributed across different layers. Specifically, the time evolution of gate field generates dynamically a positional shift of layered electrons within the 2D plane, which results in an induced Berry curvature pointing to the out-of-plane direction and enabling the intrinsic Nernst response. We reveal that the underlying intrinsic quantum geometric quantity is linked to not only the intralayer center-of-mass motion but also the interlayer coherence of layered electrons. We show that the effect can appear in nonmagnetic 2D vdW materials with chiral structural symmetry to provide the needed chirality of Hall transport, thus is dubbed as the dynamical chiral Nernst effect.

We demonstrate this effect in vdW layered structures with twisted
stacking
\cite{moireReviewEvaMacDonaldNatMater2020,moireReviewNatPhysBalents2020,moireReviewRubioNatPhys2021,moireReviewExptFolksNatRevMat2021,moireReviewJeanieLauNature2022,moireexcitonreviewNature2021,moireexcitonreviewNatRevMater2022}. In this class of materials the interlayer chiral coupling rotates the layer pseudospin about in-plane axes that are of topologically nontrivial textures in the twisted landscapes \cite{WuMacDonaldPRL2019,HongyiNSR,Zhai2020PRM,Zhai2020PRL}. The quantum layer degree of freedom activated by interlayer coherence of electronic wave functions \cite{Pesin2012,Xu2014} enables coupling the out-of-plane electric field to in-plane charge transport. As examples, we show sizable intrinsic chiral Nernst conductivities in twisted bilayer and trilayer transition metal dichalcogenides (TMDs).

\section{Quantum geometric origin of the effect}

The intrinsic contribution to Nernst effect is most readily derived within the
semiclassical theory \cite{Xiao2006,Xiao2010,Dong2020,Xiao2020EM}, which has been extended to include corrections to the band quantities due to external fields \cite{Gao2019PRL,Xiao2021OM,Xiao2021CS}. In particular, in vdW layered materials, the out-of-plane field $\boldsymbol{\mathcal{E}}=\mathcal{E}\hat{\boldsymbol{z}}$ is coupled to the
interlayer dipole moment $\hat{p}$ in the form of $-\mathcal{E}\hat{p}$, hence the temporal variation of the gate field
$\mathcal{\dot{E}}$ modifies the Bloch state
\cite{Thouless1983,Dimi2006}, inducing a \textit{k}-space
Berry connection ($\hbar\boldsymbol{k}$ is the 2D crystal momentum) \cite{Chen2023}
\begin{equation}
	\boldsymbol{\mathcal{A}}^{\mathcal{\dot{E}}}\left(  \boldsymbol{k}\right)
	=\boldsymbol{\mathcal{G}}\left(  \boldsymbol{k}\right)  \mathcal{\dot{E}}.
	\label{k-space}%
\end{equation}
For a band with index $n$, we have
\begin{equation}
	\boldsymbol{\mathcal{G}}^{n}\left(  \boldsymbol{k}\right)  =2\hbar
	^{2}\mathrm{{\operatorname{Re}}}\sum_{m\neq n}\frac{p^{nm}\left(
		\boldsymbol{k}\right)  \boldsymbol{v}^{mn}\left(  \boldsymbol{k}\right)
	}{[\varepsilon_{n}\left(  \boldsymbol{k}\right)  -\varepsilon_{m}\left(
		\boldsymbol{k}\right)  ]^{3}}, \label{BCP}%
\end{equation}
whose numerator involves the interband matrix elements of the velocity
operator and the interlayer charge dipole operator $\hat{p}$.
The specific form of $\hat{p}$ depends on the layer number \cite{Zheng2023}.
In a bilayer system, for example, $\hat{p}=ed_{0}\hat{\sigma
}_{z}$ with $\hat{\sigma}_{z}$ the Pauli matrix in the layer index
subspace and $d_{0}$ the interlayer distance, and $\varepsilon_{n}$ is the
unperturbed band energy. One can see that $\boldsymbol{\mathcal{G}}^{n}$ favors
interlayer hybridized electronic states: According to Eq.
(\ref{BCP}), if the state $|u_{n}\rangle$ is fully polarized in a specific
layer around some $\boldsymbol{k}$, then $\boldsymbol{\mathcal{G}%
}^{n}\left(  \boldsymbol{k}\right)  $ is suppressed. In addition, as
$\boldsymbol{\mathcal{G}}^{n}$ is gauge invariant, $\boldsymbol{\mathcal{A}%
}^{\mathcal{\dot{E}}}$ can be identified physically as an in-plane positional
shift of the electron \cite{Gao2014,Xiao2022MBT} induced by the time
evolution of the out-of-plane field.

This in-plane positional shift implies a dynamical generation of \textit{k}-space Berry curvature in
linear order of $\mathcal{\dot{E}}$:
\begin{equation}
	\boldsymbol{\Omega}\left(  \boldsymbol{k}\right)=
	\partial_{\boldsymbol{k}}\times\boldsymbol{\mathcal{G}}\left(  \boldsymbol{k}\right)  \mathcal{\dot{E}},
\end{equation}
which points to the out-of-plane direction. The band index $n$ has been suppressed for simplicity. $\boldsymbol{\Omega}\left(\boldsymbol{k}\right)$ will play a role in various
Berry-curvature related phenomena. Particularly, as a \textit{k}-space
effective magnetic field, it corresponds to an orbital magnetization of
topological nature that is given by the integration of $g\boldsymbol{\Omega}$
in \textit{k} space \cite{Xiao2006}, with 
%$g\left(  \varepsilon\right)  =-k_{B}T\ln[1+e^{-(\varepsilon-\mu)/k_{B}T}]$ 
\begin{equation}
	g\left(  \varepsilon\right) =-k_{B}T\ln\left[1+e^{-(\varepsilon-\mu)/k_{B}T}\right]
\end{equation}
being the grand potential density
contributed by a particular state and $k_B$ is the Boltzmann constant.

To account for the anomalous Nernst effect, it is well established that the circulating orbital
magnetization current should be discounted from the local charge current
density so that the transport current takes the form of
\cite{Xiao2006,Gao2019PRL}
\begin{equation}
	\boldsymbol{j}=e\int[d\boldsymbol{k}]f\boldsymbol{\dot{r}}-\boldsymbol{\nabla
	}\times\frac{e}{\hbar}\int[d\boldsymbol{k}]g\boldsymbol{\Omega}, \label{sum}%
\end{equation}
where $[d\boldsymbol{k}]$ is shorthand for $\sum_{n}d^{2}\boldsymbol{k}%
/(2\pi)^{2}$. One observes that the second term arises from the
topological orbital magnetization due to the \textit{k}-space Berry curvature.
On the other hand, the first term is the conventional expression for the
charge current, with $\boldsymbol{\dot{r}}$ being the semiclassical velocity
of an electron wave packet and $f\left(  \varepsilon\right)  $ the equilibrium
Fermi-Dirac distribution function. We do not consider the off-equilibrium
distribution function $\sim\boldsymbol{\nabla}T$ in the first order of
in-plane temperature gradient, because it does not contribute to the Nernst
current in nonmagnetic systems --the present focus-- as required by
time reversal symmetry. Moreover, since the temperature gradient $\boldsymbol{\nabla}T$ does not
enter into $\boldsymbol{\dot{r}}$, the first term in Eq. (\ref{sum}) does not
give any current in the linear order of $\boldsymbol{\nabla}T$. As a result, the Nernst current driven by $\boldsymbol{\nabla}T$ arises totally from the second term
of Eq. (\ref{sum}), which reads
\begin{equation}
	\boldsymbol{j}_{\text{Nernst}}=\vartheta\boldsymbol{\mathcal{\dot{E}}}\times\boldsymbol{\nabla
	}T, \label{Hall}%
\end{equation}
where the response coefficient%
\begin{equation}
	\vartheta=e\int[d\boldsymbol{k}]\frac{\partial f\left(  \varepsilon\right)
	}{\partial\varepsilon}\frac{\varepsilon-\mu}{T}[\boldsymbol{v}\times
	\boldsymbol{\mathcal{G}}\left(  \boldsymbol{k}\right)  ]_{z} \label{theta}
\end{equation}
is intrinsic to the band structure and is a Fermi surface property.

One finds that $\vartheta$ is a time-reversal even pseudoscalar, thus it is
invariant under rotation, but flips sign under space inversion, reflection and
rotoreflection symmetries. As such, $\vartheta$ is allowed only if the
system possesses a chiral crystal symmetry~\cite{ZhaiLayerHall2022,Chen2023}. This new type of Nernst response, dubbed as the dynamical chiral Nernst effect,
is the first scenario of intrinsic Nernst response without magnetic degree of freedom. 2D chiral lattices, such as Tellurene \cite{Tellurene2018,Tellurene2023} and twisted layers of honeycomb lattice ~\cite{Gao2020,ZhaiLayerHall2022,Chen2023}, are suitable platforms for this effect. In particular, in twisted structures, the chirality is locked to the twist direction: Since twisted
configurations with opposite twist angles are mirror images of each other, whereas
the mirror reflection flips the sign of $\vartheta$, the direction of the Nernst
current is reversed by reversing the twist direction. This character renders a unique knob to control the effect.

For a sine AC gate field $\mathcal{E}=\mathcal{E}_{0}\sin\omega t$,
the oscillating Nernst current is given by
\begin{equation}
	\boldsymbol{j}_{\text{Nernst}}=\alpha_{\text{H}}\cos\omega t\,\boldsymbol{\hat{z}}%
	\times\boldsymbol{\nabla}T~~\text{with}~~\alpha_{\text{H}}=\omega\mathcal{E}_{0}\vartheta.
\end{equation}
Here $\omega$ is required to be below the threshold for direct interband
transition in order to validate the semiclassical treatment, and
$\alpha_{\text{H}}$ has the dimension of Nernst conductivity and quantifies
the Nernst response with respect to the in-plane temperature gradient.

%%%%%%%%%%%%%%%%%%%%%%%%%%%%%%%%%%%%%%%%%%%%%%%%%%%%%%%%%%%%%%%%%%%%%%%%%
%%%%%%%%%%%%%%%%%%%%%%%%%%%%%%%%%%%%%%%%%%%%%%%%%%%%%%%%%%%%%%%%%%%%%%%%%

Next we study the effect
quantitatively by considering twisted bilayer and trilayer TMDs, using MoTe$_{2}$ as an
example~\cite{WuMacDonaldPRL2019,HongyiNSR}. Figure~\ref{Fig:bilayer}(a) shows the schematics of the experimental setup. We take
$\omega/2\pi=0.1$~THz and $\mathcal{E}_{0}d_{0}=10$~mV
\cite{moireReviewEvaMacDonaldNatMater2020,moireexcitonreviewNature2021,moireexcitonreviewNatRevMater2022}
in what follows.

%%%%%%%%%%%%%%%%%%%%%%%%

\section{Application to twisted homobilayer TMDs}

We employ the four-band model of twisted bilayer MoTe$_{2}$ in Ref.~\cite{WuMacDonaldPRL2019}. The Hamiltonian for spin-up carriers in the $K$ valley is
\begin{equation}
	H_{\text{bilayer}}=
	\left(
	\setlength{\arraycolsep}{1pt}
	\begin{matrix}
		\mathcal{T}_t+\Delta_t&\Delta_T^\dagger(\boldsymbol{\delta})\\
		\Delta_T(\boldsymbol{\delta})&\mathcal{T}_b+\Delta_b
	\end{matrix}
	\right),
	\label{eq:bilayer hamil}
\end{equation}
where the subscripts $t$ and $b$ refer to the top and bottom layer, $\boldsymbol{\delta}=\theta\hat{z}\times\boldsymbol{r}$ is the local displacement vector at position $\boldsymbol{r}$ between the two layers.  $\mathcal{T}_l$ with $l=t,\,b$ is the layer-dependent kinetic energy
\begin{equation}
	\mathcal{T}_l=e^{-il\frac{\theta}{4}\xi_z}[\hbar v_F(\boldsymbol{k}-\boldsymbol{\kappa}_l)\cdot\boldsymbol{\xi}]e^{+il\frac{\theta}{4}\xi_z}+{\rm diag}(\Delta_g,0),\label{eq:kinetic}
\end{equation}
in which $l=1$ for $t$ and $l=-1$ for $b$ in the exponential terms, and $\Delta_g=1.1$ eV is the energy gap between conduction and valence bands of monolayer MoTe$_2$. $\Delta_T$ is the interlayer tunneling potential
\begin{equation}
	\Delta_T(\boldsymbol{\delta})=U_0+U_1e^{-i\boldsymbol{G}_2\cdot\boldsymbol{\delta}}+U_2e^{-i\boldsymbol{G}_3\cdot\boldsymbol{\delta}},
	\label{eq:InterlayerTunneling}
\end{equation}
where $U_{n=0,1,2}=\begin{pmatrix}
w_{cc} & w_{cv}e^{-i\frac{2\pi}{3}n} \\w_{vc}e^{+i\frac{2\pi}{3}n}&w_{vv}
\end{pmatrix}$, $\boldsymbol{G}_i$ is generated from the counterclockwise $(i-1)\pi/3$ rotation of $\boldsymbol{G}_1=(0,\,1)4\pi/(\sqrt{3}a_0)$, $a_0=3.47$ \AA~is the lattice constant of monolayer MoTe$_2$, and $w_{cc}=-2$ meV, $w_{vv}=-8.5$ meV, $w_{cv}=w_{vc}=15.3$ meV are the tunneling parameters (subscripts $c$ and $v$ represent conduction and valence bands, respectively). $\Delta_l$ is the layer-dependent electrostatic potential with
\begin{equation}
	\Delta_t
	=-\Delta_b
	=\sum_{j=1,3,5}\sin(\boldsymbol{G}_j\cdot\boldsymbol{\delta})\left(
	\begin{matrix}
		V_c&0\\
		0&V_v
	\end{matrix}
	\right),\label{eq:Electrostatic}
\end{equation}
where $V_c=-11.94$ meV and $V_v=-16$ meV.

%%%%%%%%%%%%%%%%%%%%%%%%%%%%%%%%%%%%%%%%%%%%

\begin{figure}[t]
	\centering
	\includegraphics[width=3.4in]{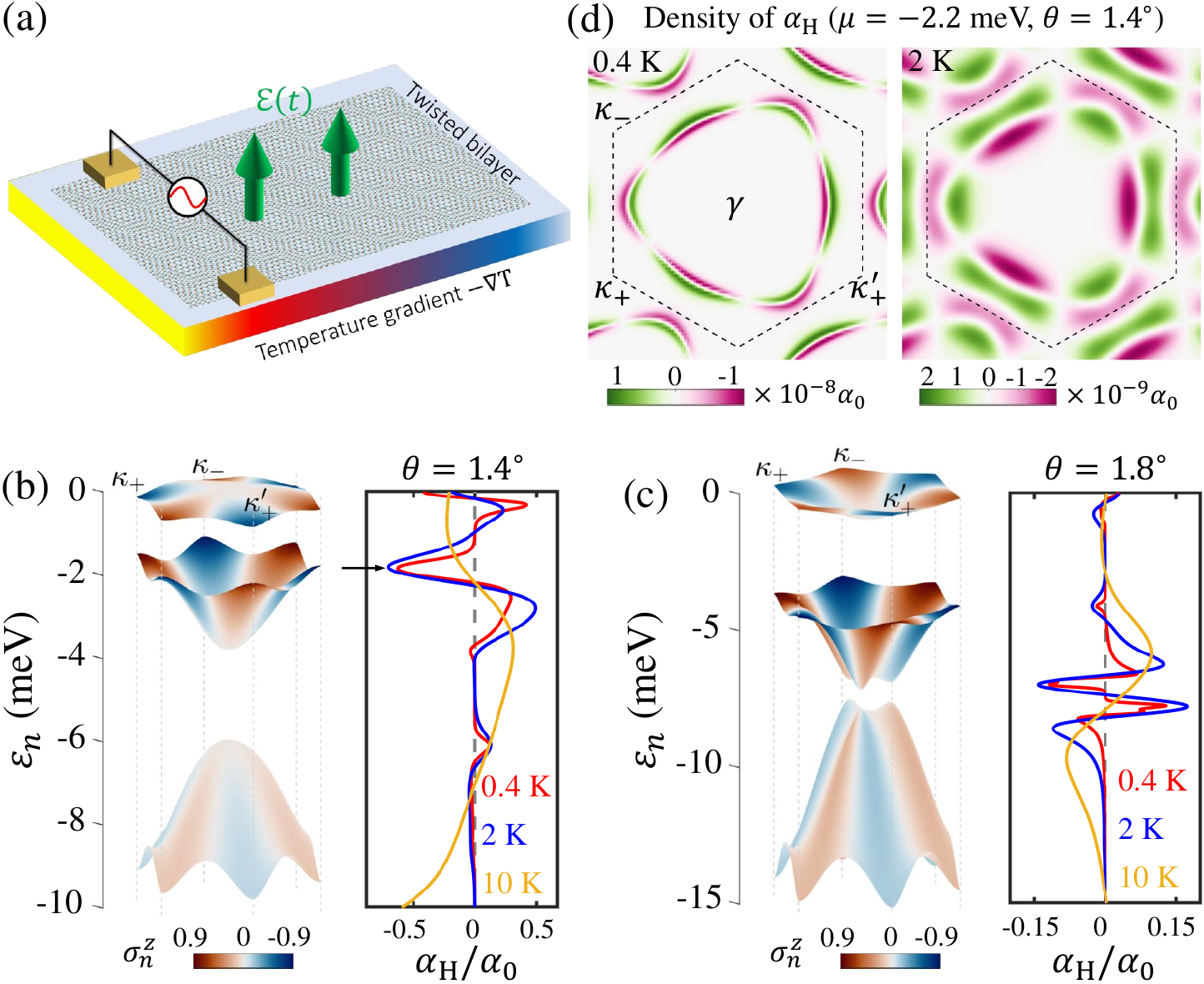}
	\caption{(a) Schematics of the experimental setup for the dynamical chiral Nernst effect. (b, c) Energy bands and Nernst conductivity $\alpha_{\rm H}$ at various temperatures from the $K$ valley at twist angle of 1.4$^\circ$ (b) and 1.8$^\circ$ (c). $\alpha_0=ek_B/h\approx3.34$ {\rm nA/K}. (d) The density of Nernst conductivity in the momentum space at two different temperatures when the Fermi level is $-2.2$ meV [black arrow in (b)] and $\theta=1.4^\circ$.}
	\label{Fig:bilayer}
\end{figure}

%%%%%%%%%%%%%%%%%%%%%%%%%%%%%%%%%%%%%%%%%%%%

Fig.~\ref{Fig:bilayer}(b) shows the valence band structure, and the Nernst conductivity $\alpha_{\rm H}$ measured in units of $\alpha_0=ek_B/h\approx3.34$ {\rm nA/K}~\cite{checkelsky2009thermopower,sharma2016nernst,dau2019valley}, in the $K$ valley of $1.4^\circ$ twisted bilayer MoTe$_2$. The color on the band structures denotes $\sigma^{z}_{n}=\braket{u_n|\hat{\sigma}_{z}|u_n}$, which characterizes the strength of interlayer hybridization~\cite{ZhaiLayerHall2022,Chen2023}. Specifically, $\sigma^{z}_{n}\sim0$ ($|\sigma^{z}_{n}|\sim1$) indicates strong (weak) interlayer hybridization. Eq.~(\ref{BCP}) dictates that a large $\bCalG$, thus $\alpha_{\text{H}}$, requires strong interlayer hybridization and small interband energy separation. In Fig.~\ref{Fig:bilayer}(b), the first two energy bands are close to each other, and interlayer hybridization is strong except around the Brillouin zone corners. Consequently, one observes large conductivity peaks at low temperatures within the energy window of these two bands [red and blue curves, Fig.~\ref{Fig:bilayer}(b)]. In contrast, the third energy band contributes much smaller conductivity due to the large separation to other bands, although the states are also strongly interlayer hybridized.

The conductivity has a strong dependence on the temperature. At low temperatures, $\partial_{\varepsilon} f/T$ in the integrand of Eq.~(\ref{theta}) is sharply peaked with large magnitude around the Fermi level, while it becomes broadened with reduced magnitude when the temperature increases.
To illustrate this temperature modulation, Fig.~\ref{Fig:bilayer}(d) presents the density of $\alpha_{\rm H}$ in the momentum space at 0.4 K and 2 K when the Fermi level is at $-2.2$ meV [black arrow in Fig.~\ref{Fig:bilayer}(b)]. Clearly, the density becomes more widely distributed as the temperature increases, which result in broadened conductivity peaks [red vs blue curves, Fig.~\ref{Fig:bilayer}(b)]. For temperatures $\mathcal{O}(10)$ K, the conductivity remains observable and exhibits strong thermal smearing features [yellow curve, Fig.~\ref{Fig:bilayer}(b)].

Fig.~\ref{Fig:bilayer}(c) shows the results of twisted bilayer MoTe$_2$ at another twist angle of $1.8^\circ$. As the second energy band moves closer to the third, the most pronounced conductivity peaks now locate in the energy window that covers these two bands. All the other features remain qualitatively the same as those of $1.4^\circ$.

%%%%%%%%%%%%%%%%%%%%%%%%%%%%%%%%%%%%%%%%%%%%%%%%%%%%%%%%%%%%%%%%%%%%%%%%%%%%%%%%
%%%%%%%%%%%%%%%%%%%%%%%%%%%%%%%%%%%%%%%%%%%%%%%%%%%%%%%%%%%%%%%%%%%%%%%%%%%%%%%%

\section{Application to twisted homotrilayer TMDs}
\begin{figure*}[t]
	\centering
	\includegraphics[width=5in]{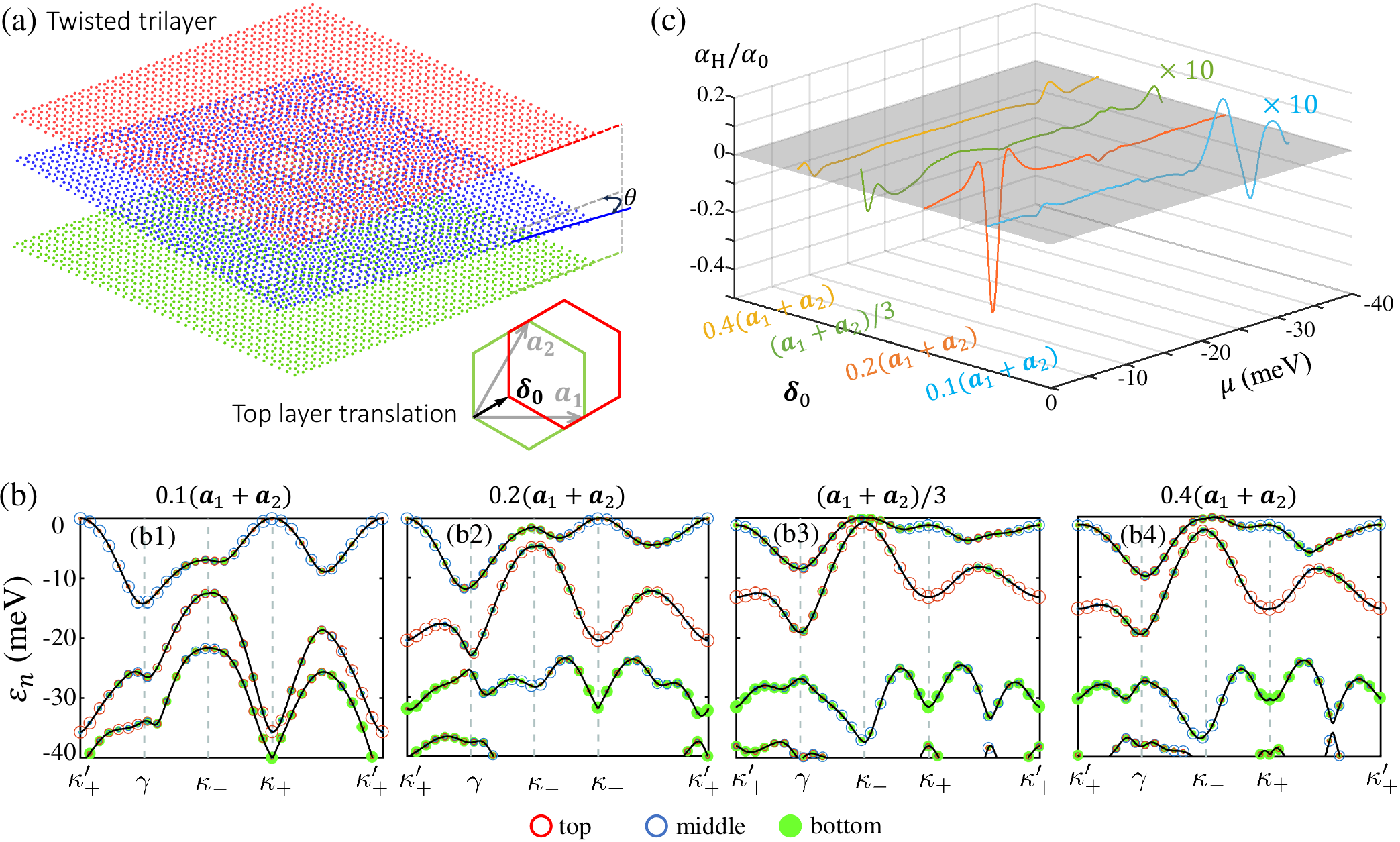}
	\caption{(a) Schematics of twisted homotrilayer TMDs. The top (red) and bottom (green) layers are parallel, while the middle (blue) layer is misoriented by $\theta$. The inset shows the global displacement $\boldsymbol{\delta}_0$ between the top and bottom layers, which is parameterized by $\boldsymbol{a}_1=a_0(1,0)$ and $\boldsymbol{a}_2=a_0(1/2,\sqrt{3}/2)$. (b) The valence bands from the $K$ valley of $\theta=3.5^\circ$ twisted trilayer MoTe$_2$ with different top layer translations: (from left to right) $\boldsymbol{\delta}_0=0.1(\boldsymbol{a}_1+\boldsymbol{a}_2)$, $\boldsymbol{\delta}_0=0.2(\boldsymbol{a}_1+\boldsymbol{a}_2)$, $\boldsymbol{\delta}_0=(\boldsymbol{a}_1+\boldsymbol{a}_2)/3$, and $\boldsymbol{\delta}_0=0.4(\boldsymbol{a}_1+\boldsymbol{a}_2)$. The red, blue, and green circles denote the percentage of each layer in the Bloch states (large radius corresponds to high percentage). (c) The chiral Nernst conductivity at $T=4$ K associated with the various cases in (b).}
	\label{Fig:trilayer}
\end{figure*}

Next we consider twisted homotrilayer TMDs. Specifically, we assume that the top and bottom layers are parallel, while the middle layer is misorientated by $\theta$. We also assume that the top layer can be translated with respect to the bottom layer by $\boldsymbol{\delta}_0$ from the fully aligned configuration [see Fig.~\ref{Fig:trilayer}(a)]. Such a trilayer geometry shares the same moir\'{e} period with the bilayer case.

The Hamiltonian for the twisted trilayer can be built from that of the bilayer \cite{Zheng2023}:
\begin{equation}
	H_{\text{trilayer}}=
	\left(
	\setlength{\arraycolsep}{1pt}
	\begin{matrix}
		\mathcal{T}_t+Q_t&\Delta_T^\dagger(\boldsymbol{\delta}^\prime)&0\\
		\Delta_T(\boldsymbol{\delta}^\prime)&\mathcal{T}_m+Q_m&\Delta_T(\boldsymbol{\delta})\\
		0&\Delta_T^\dagger(\boldsymbol{\delta})&\mathcal{T}_b+Q_b
	\end{matrix}
	\right),
	\label{eq:trilayer hamil}
\end{equation}
where the subscripts $t$, $m$ and $b$ refer to the top, middle and bottom layer, $\boldsymbol{\delta}^\prime=\theta\hat{z}\times\boldsymbol{r}+\boldsymbol{\delta}_0$ $(\boldsymbol{\delta}=\theta\hat{z}\times\boldsymbol{r}$) is the local displacement vector at position $\boldsymbol{r}$ between the top-middle (middle-bottom) layer. $\mathcal{T}_l$ with $l=t,\,m,\,b$ is given in Eq.~(\ref{eq:kinetic}), in which $l=+1$ for $t$ and $b$, and $l=-1$ for $m$ in the exponential terms. $\Delta_T$ is given by Eq.~(\ref{eq:InterlayerTunneling}).
The layer-dependent electrostatic potential $Q_l$ is built with Eq.~(\ref{eq:Electrostatic}) as~\cite{Qingjun2DM}:
\begin{equation}
	\begin{split}
		&Q_t=-Q_b=[\Delta_t(\boldsymbol{\delta}^\prime)-\Delta_t(\boldsymbol{\delta})],\\
		&Q_m=-[\Delta_t(\boldsymbol{\delta}^\prime)+\Delta_t(\boldsymbol{\delta})].
	\end{split}
\end{equation}

In addition, to evaluate the Nernst conductivity, the interlayer charge dipole operator in Eq.~(\ref{BCP}) is replaced by $\hat{p}=ed_0\,\text{diag}(1,\,0,\,-1)$ in the trilayer geometry~\cite{Zheng2023}.

%%%%%%%%%%%%%%%%%%%%%%%%%%%%%%%%%%%%%%%%%%%%%%%%%%%%%%%%%%%%%

%%%%%%%%%%%%%%%%%%%%%%%%%%%%%%%%%%%%%%%%%%%%%%%%%%%%%%%%%%%%%

When the top and bottom layers are aligned with $\boldsymbol{\delta}_0=0$, the structure has mirror symmetry in the out-of-plane direction, thus the chiral Nernst effect is forbidden. As the top layer is translated, the mirror symmetry is broken and the Nernst effect is turned on.
Fig.~\ref{Fig:trilayer}(b) shows the modulation of the valence bands and layer hybridization by $\boldsymbol{\delta}_0$ in the $K$ valley of $3.5^\circ$ twisted trilayer MoTe$_2$. For example, when $\boldsymbol{\delta}_0$ is tuned from $0.1(\boldsymbol{a}_1+\boldsymbol{a}_2)$ to $0.2(\boldsymbol{a}_1+\boldsymbol{a}_2)$ [see Fig.~\ref{Fig:trilayer}(a) inset for $\boldsymbol{a}_1$ and $\boldsymbol{a}_2$], the second valence band moves closer to the first, and the energy separations between the second and third bands are enlarged in general [Fig.~\ref{Fig:trilayer}(b1) vs (b2)]. For larger $\boldsymbol{\delta}_0$, the separation between the first two bands can be even smaller at certain locations of the momentum space [Figs.~\ref{Fig:trilayer}(b3) and (b4)].
When $\boldsymbol{\delta}_0$ is small, one notices that the low-energy states are strongly polarized to the middle layer, see e.g., the dominant blue circles for $|\varepsilon_{n}|<5$ meV in Fig.~\ref{Fig:trilayer}(b1). As the translation increases, the low-energy states become strongly layer hybridized, as shown in Figs.~\ref{Fig:trilayer}(b2)--(b4).
Such modulations of energy and layer hybridization affect the Nernst conductivity.
For example, the large energy separations and weak layer hybridization at low energies in Fig.~\ref{Fig:trilayer}(b1) dictates that the corresponding conductivity is negligible, which is confirmed by the cyan curve in Fig.~\ref{Fig:trilayer}(c).
In contrast, remarkably strong conductivity peaks can be found with a Fermi level around $10$ meV for the intermediate translation of $\boldsymbol{\delta}_0=0.2(\boldsymbol{a}_1+\boldsymbol{a}_2)$ [orange curve, Fig.~\ref{Fig:trilayer}(c)], and large conductivity peaks can be achieved with very low doping near the topmost band edges for large translations of $\boldsymbol{\delta}_0=(\boldsymbol{a}_1+\boldsymbol{a}_2)/3$ and $\boldsymbol{\delta}_0=0.4(\boldsymbol{a}_1+\boldsymbol{a}_2)$ [green and yellow curves, Fig.~\ref{Fig:trilayer}(c)].

\section{Discussion}
We have proposed a novel effect, the dynamical chiral Nernst effect, which is unique to 2D vdW materials with chiral lattice structures. This is also the first Nernst response of an intrinsic nature determined solely by the band structures in nonmagnetic materials. We revealed that its quantum geometric origin is rooted in the interlayer coherence of electronic states endowed by strong interlayer quantum tunneling. The effect is shown to be sizable in typical twisted bilayer and trilayer TMDs, and can be feasibly tuned by twist angle and interlayer translation.

Compared to the intrinsic anomalous Nernst effect in magnetic systems caused by \textit{k}-space Berry curvature, the intrinsic chiral Nernst effect in our work does not require magnetic field or internal magnetization, and shows similar or even larger magnitudes that are well within experimental capacity. Our calculations showed that the chiral Nernst conductivity can reach $\sim 0.5 \alpha_0$ in twisted bilayer TMDs [i.e., 2.4 A/(K$\cdot$ m) with the thickness of bilayer MoTe$_2$] and trilayer TMDs. In comparison, the intrinsic anomalous Nernst conductivity can reach $\sim3\times10^{-4}\alpha_0$ in monolayer TMDs placed on a magnetic insulating substrate~\cite{yu2015thermally}, and the experiment on magnetic topological semimetal Co$_2$MnGa reported anomalous Nernst conductivities around 3 A/(K$\cdot$ m) \cite{Sakai2018}.

This work opens a new route towards in-plane thermoelectric conversion by out-of-plane
dynamical control of layered vdW structures \cite{zhai2022ultrafast}. The
study can be generalized to chiral thermal Hall effects and thermoelectric Hall transport of spin and valley degrees of freedom.

\section{Acknowledgments}

This work is supported by the National Key R\&D Program of China (Grant No. 2020YFA0309600), the Research Grant Council of Hong Kong (AoE/P-701/20, HKU SRFS2122-7S05, A-HKU705/21), and New Cornerstone Science Foundation. C.X. also acknowledges support by the UM Start-up Grant (SRG2023-00033-IAPME).

\bibliographystyle{apsrev4-1}
\bibliography{ChiralNernstRef}

\end{document}